\documentclass{article}
\usepackage[psamsfonts]{amssymb}
\usepackage{amsmath}
\usepackage{cite}
\usepackage{eufrak}
\def\e{\EuFrak{e}}

\def\d{\mathrm{d}}
\def\dim{\mathrm{dim}}
\def\id{\mathbf{1}}
\def\reg{\mathrm{reg}}
\def\c{\star}
\def\ir{\mathrm{i}}
\def\vo{\mathrm{vol}(G)}
\begin{document}
\begin{titlepage}
\noindent{\large\textbf{Field theories on spaces with linear
fuzziness}}

\vskip 2 cm

\begin{center}{Amir~H.~Fatollahi{\footnote
{ahfatol@gmail.com}} \& Mohammad~Khorrami{\footnote
{mamwad@mailaps.org}} } \vskip 5 mm \textit{ Department of
Physics, Alzahra University,
             Tehran 1993891167, Iran. }

\end{center}

\begin{abstract}
\noindent A noncommutative space is considered the position
operators of which satisfy the commutativity relations of a Lie
algebra. The basic tools for calculation on this space, including
the product of the fields, inner product and the proper measure
for integration are derived. Some general aspects of perturbative
field theory calculations on this space are also discussed. Among
the features of such models is that they are free from
ultraviolet divergences (and hence free from UV/IR mixing as
well), if the group is compact. The example of the group SO(3) or
SU(2) is investigated in more detail.
\end{abstract}
\end{titlepage}
\section{Introduction}
During recent years much attention has been paid to the
formulation and study of field theories on noncommutative spaces.
The motivation is the natural appearance of noncommutative spaces
in some areas of physics, for example recently in the string
theory. In particular it has been understood that the longitudinal
directions of D-branes in the presence of a constant B-field
background appear to be noncommutative, as seen by the ends of
open strings \cite{9908142,99-2,99-3,99-4}. In this case the
coordinates satisfy the canonical relation
\begin{equation}\label{f.1}
[\widehat x_a,\widehat x_b]=\ir\,\theta_{a\, b}\,\id,
\end{equation}
in which $\theta$ is an antisymmetric constant tensor and $\id$
represents the unit operator. The theoretical and phenomenological
implications of possible noncommutative coordinates have been
extensively studied; see \cite{reviewnc}.

In the present paper the case beyond the canonical one is
investigated. In particular a model is considered in which the
(dimensionless) spatial positions operators satisfy the
commutation relations of a Lie algebra \cite{wess}:
\begin{equation}\label{f.2}
[\widehat x_a,\widehat x_b]=f^c{}_{a\, b}\,\widehat x_c,
\end{equation}
where $f^c{}_{a\,b}$'s are structure constants of a Lie algebra.

An example of this kind is the algebra SO(3), or SU(2).
A special case of this is the so called fuzzy sphere \cite{madore},
where an irreducible representation of the position operators is
used which makes the Casimir of the algebra, $(\hat x_1)^2+(\hat x_2)^2+(\hat x_3)^2$,
a multiple of the identity operator (a constant, hence the name
sphere). One can consider the square root of this Casimir as the
radius of the fuzzy sphere. This is, however, a noncommutative
version of a two-dimensional space (sphere).

In the present work a model is introduced in which the noncommutativity
is again taken to be that of a group, but no specific irreducible representation
is considered. In particular, we employ the regular representation of the group,
which contains all representations. As a consequence and for the special case
of SU(2) group, in our model one is dealing with the whole of 3-dimensional space,
instead of a 2-dimensional subspace of it as in fuzzy sphere case.
The space of the corresponding momenta is an ordinary (commutative)
space, and is compact iff the group is compact. In fact one can
consider the momenta as the coordinates of the group. So a by-product
of such a model would be the elimination of any ultraviolet divergence
in any field theory constructed on such a space. One important
implication of the elimination of the ultraviolet divergences, as
we go in more detail later, would be that there will not remain any place
for the so called UV/IR mixing effect, which is known as a
common artifact one expects to face with in the models with
canonical noncommutativity, the algebra (\ref{f.1}).

Here we consider the noncommutativity only among spatial coordinates.
In \cite{majid,ruegg,amelino} a situation is considered in which
noncommutativity is introduced between spatial directions and
time, that is
\begin{align}\label{f.3}
[\widehat x_a,\widehat t\,]=&\,\frac{\ir}{\kappa}\,\widehat x_a,\cr
[\widehat x_a,\widehat x_b]=&\,0,
\end{align}
where $\kappa$ is a constant.

The scheme of this paper is the following. In section 2, some
basic aspects of the group algebra are reviewed, mainly to fix
notations. In section 3 a model is investigated containing a real
field with momenta in a compact group. In section 4 this case is
specialized to the group SU(2) or SO(3). Section 5 is devoted to
concluding remarks, and a discussion of the possible divergences
of the theories is presented.
\section{The group algebra}
For a compact group $G$, there is a unique measure $\d U$ (up to a
multiplicative constant) with the invariance properties
\begin{align}\label{f.4}
\d (V\,U)&=\d U,\nonumber\\
\d (U\,V)&=\d U,\nonumber\\
\d (U^{-1})&=\d U,
\end{align}
for any arbitrary element ($V$) of the group. These mean that this
measure is invariant under the left-translation,
right-translation, and inversion. This measure, the
(left-right-invariant) Haar measure, is unique up to a
normalization constant, which defines the volume of the group:
\begin{equation}\label{f.5}
\int_G\d U=\vo.
\end{equation}
Using this measure, one constructs a vector space as follows.
Corresponding to each group element $U$ an element $\e(U)$ is
introduced, and the elements of the vector space are linear
combinations of these elements:
\begin{equation}\label{f.6}
f:=\int\d U\;f(U)\,\e(U),
\end{equation}
The group algebra is this vector space, equipped with the
multiplication
\begin{equation}\label{f.7}
f\,g:=\int\d U\,\d V\; f(U)\,g(V)\,\e(U\,V),
\end{equation}
where $(U\,V)$ is the usual product of the group elements. $f(U)$
and $g(U)$ belong to a field (here the field of complex numbers).
It can be seen that if one takes the central extension of the
group U(1)$\times\cdots\times$U(1), the so-called Heisenberg
group, with the algebra (\ref{f.1}), the above definition results
in the well-known star product of two functions, provided $f$ and
$g$ are interpreted as the Fourier transforms of the functions.

So there is a correspondence between functionals defined on the
group, and the group algebra. The definition (\ref{f.7}) can be
rewritten as
\begin{align}\label{f.8}
(f\,g)(W)=&\int\d V\;f(W\,V^{-1})\,g(V),\nonumber\\
=&\int\d U\;f(U)\,g(U^{-1}\,W).
\end{align}

Using the Schur's lemmas, one proves the so called grand
orthogonality theorem which states that there is an orthogonality
relation between the matrix functions of the group:
\begin{equation}\label{f.9}
\int\d U\; U_\lambda{}^a{}_b\,U^{-1}_\mu{}^c{}_d=
\frac{\vo}{\dim_\lambda}\,\delta_{\lambda\,\mu}\,\delta^a_d\,\delta^c_b,
\end{equation}
where $U_\lambda$ is the matrix of the element $U$ of the group in
the irreducible representation $\lambda$, and $\dim_\lambda$ is
the dimension of the representation $\lambda$. Exploiting the
unitarity of these representations, one can write (\ref{f.9}) in
the more familiar form
\begin{equation}\label{f.10}
\int\d U\; U_\lambda{}^a{}_b\,U^*_\mu{}_d{}^c=
\frac{\vo}{\dim_\lambda}\,\delta_{\lambda\,\mu}\,\delta^a_d\,\delta^c_b.
\end{equation}
Using this orthogonality relation, one can obtain an orthogonality
relation between the characters of the group:
\begin{equation}\label{f.11}
\int\d
U\;\chi_\lambda(U)\,\chi_\mu(U^{-1})=\vo\,\delta_{\lambda\,\mu},
\end{equation}
or
\begin{equation}\label{f.12}
\int\d
U\;\chi_\lambda(U)\,\chi^*_\mu(U)=\vo\,\delta_{\lambda\,\mu},
\end{equation}
where
\begin{equation}\label{f.13}
\chi_\lambda(U):=U_\lambda{}^a{}_a.
\end{equation}

The delta distribution is defined through
\begin{equation}\label{f.14}
\int\d U\;\delta(U)\,f(U):=f(\id),
\end{equation}
where $\id$ is the identity element of the group. It is easy to
see that this delta distribution is invariant under similarity
transformations, as well as inversion of the argument:
\begin{align}\label{f.15}
\delta(V\,U\,V^{-1})&=\delta(U),\cr \delta(U^{-1})&=\delta(U).
\end{align}

The regular representation of the group is defined through
\begin{equation}\label{f.16}
U_\reg\,\e(V):=\e(U\,V),
\end{equation}
from which it is seen that the matrix element of this linear
operator is
\begin{equation}\label{f.17}
U_\reg(W,V)=\delta(W^{-1}\,U\,V).
\end{equation}
This shows that the trace of the regular representation is
proportional to the delta distribution:
\begin{align}\label{f.18}
\chi_\reg(U)=&\int\d V\;U_\reg(V,V),\cr =&\vo\,\delta(U).
\end{align}
So the delta distribution can be expanded in terms of the matrix
functions (in fact in terms of the characters of irreducible
representations). The result is
\begin{equation}\label{f.19}
\delta(U)=\sum_\lambda\frac{\dim_\lambda}{\vo}\,\chi_\lambda(U),
\end{equation}
or
\begin{align}\label{f.20}
\delta(U\,V^{-1})=&\sum_\lambda\frac{\dim_\lambda}{\vo}\,
U_\lambda{}^a{}_b\,V^{-1}_\lambda{}^b{}_a,\cr
=&\sum_\lambda\frac{\dim_\lambda}{\vo}\,U_\lambda{}^a{}_b\,V^*_\lambda{}_a{}^b.
\end{align}
This shows that other functions are also expandable in terms of
the matrix functions:
\begin{equation}\label{f.21}
f(U)=\sum_\lambda\frac{\dim_\lambda}{\vo}\,
U_\lambda{}^a{}_b\,f_\lambda{}_a{}^b,
\end{equation}
where
\begin{align}\label{f.22}
f_\lambda{}_a{}^b:=&\int\d V\;V^{-1}_\lambda{}^b{}_a\,f(V),\cr
=&\int\d V\;V^*_\lambda{}_a{}^b\,f(V).
\end{align}
Using this and (\ref{f.8}), one arrives at
\begin{equation}\label{f.23}
(f\,g)_\lambda{}_a{}^b=f_\lambda{}_a{}^c\,g_\lambda{}_c{}^b.
\end{equation}

Next, one can define an inner product on the group algebra.
Defining
\begin{equation}\label{f.24}
\langle\e(U),\e(V)\rangle:=\delta(U^{-1}\,V),
\end{equation}
and demanding that the inner product be linear with respect to its
second argument and antilinear with respect to its first argument,
one arrives at
\begin{align}\label{f.25}
\langle f,g\rangle=&\int\d U\;f^*(U)\,g(U),\cr
=&\sum_\lambda\frac{\dim_\lambda}{\vo}\,f^*_\lambda{}^a{}_b\,g_\lambda{}_a{}^b.
\end{align}

Finally, one defines a star operation through
\begin{equation}\label{f.26}
f^\c(U):=f^*(U^{-1}).
\end{equation}
This is in fact equivalent to definition of the star operation in
the group algebra as
\begin{equation}\label{f.27}
[\e(U)]^\c:=\e(U^{-1}).
\end{equation}
It is then easy to see that
\begin{align}\label{f.28}
(f\,g)^\c=&g^\c\,f^\c,\\ \label{f.29} \langle f, g\rangle=&
(f^\c\,g)(\id).
\end{align}
Here a note is in order. While the results of this section were
obtained for compact groups, in some cases the compactness is not
necessary. It is easy to see that provided (\ref{f.4}) holds,
(\ref{f.6}) to (\ref{f.8}), (\ref{f.14}) to (\ref{f.17}),
(\ref{f.24}), the first equality in (\ref{f.25}), and (\ref{f.26})
to (\ref{f.29}) are still true, even if the group is noncompact.
\section{The real scalar field}
To give motivation for the particular form of the action which is
going to be written for a real scalar field, let's first consider
the real scalar field on an ordinary $\mathbb{R}^D$ space.
\subsection{The real scalar field: the Fourier transform picture}
To be consistent with the notation used throughout this paper, the
Fourier transform (only on space) of the field is denoted by
$\phi$, while the field itself is denoted by $\tilde\phi$. So,
\begin{equation}\label{f.30}
\tilde\phi(\mathbf{r})=\int\frac{\d^D
k}{(2\,\pi)^D}\;\phi(\mathbf{k})\,\exp(\ir\,\mathbf{r}\cdot\mathbf{k}).
\end{equation}
An action for a scalar field is
\begin{equation}\label{f.31}
S=\int\d t\,\d^D
r\;\left\{\frac{1}{2}\,\left[\dot{\tilde\phi}(\mathbf{r})\,
\dot{\tilde\phi}(\mathbf{r})+ \tilde\phi(\mathbf{r})\,\tilde
O(\nabla)\,\tilde\phi(\mathbf{r})\right]-\sum_{j=3}^n\frac{g_j}{j!}\,
[\tilde\phi(\mathbf{r})]^j\right\},
\end{equation}
where $g_j$'s are constants and $\tilde O(\nabla)$ is a
differential operator. This action is translation-invariant, that
is invariant under transformations
\begin{equation}\label{f.32}
\tilde\phi(\mathbf{r})\to{\tilde\phi}'(\mathbf{r}):=
{\tilde\phi}'(\mathbf{r}-\mathbf{a}),
\end{equation}
where $\mathbf{a}$ is constant.

One can write the action (\ref{f.31}) and the transformation
(\ref{f.32}) in terms of the Fourier transforms:
\begin{align}\label{f.33}
S=\int\d t\;\Bigg\{\frac{1}{2}&\int\frac{\d^D k_1\,\d^D
k_2}{(2\,\pi)^{2\,D}}\;[\dot\phi(\mathbf{k}_1)\,
\dot\phi(\mathbf{k}_2)\cr +&
\phi(\mathbf{k}_1)\,O(\mathbf{k}_2)\,\phi(\mathbf{k}_2)]\,
[(2\,\pi)^D\,\delta(\mathbf{k}_1+\mathbf{k}_2)]\cr
-&\sum_{j=3}^n\frac{g_j}{j!}\int\left[\prod_{l=1}^j\frac{\d^D
k_l\;\phi(\mathbf{k}_l)}{(2\,\pi)^D}\right]\;
[(2\,\pi)^D\,\delta(\mathbf{k}_1+\cdots+\mathbf{k}_j)]\Bigg\},
\end{align}
and
\begin{equation}\label{f.34}
\phi(\mathbf{k})\to\phi'(\mathbf{k}):=
\exp(-\ir\,\mathbf{k}\cdot\mathbf{a})\,\phi(\mathbf{k}).
\end{equation}

Considering the space of $\mathbf{k}$'s as a group
($\mathbb{R}^D$), one notices that $(\d^D k)/(2\,\pi)^D$ is the
measure of this group which is invariant under right translation,
left translation, and inversion. It is not normalizable in the
sense (\ref{f.5}), as this group is not compact. One also notices
that $\exp(-\ir\,\mathbf{k}\cdot\mathbf{a})$ is nothing but the
representation $\mathbf{a}$ of the group element corresponding to
the coordinates $\mathbf{k}$. As this representation is one
dimensional, $\exp(-\ir\,\mathbf{k}\cdot\mathbf{a})$ is also the
determinant of this representation.
\subsection{The real scalar field on general compact groups}
A real scalar field $\phi$ is defined as a real member of the
group algebra:
\begin{equation}\label{f.35}
\phi^\c=\phi.
\end{equation}
A simple action for this field can be of the form
\begin{equation}\label{f.36}
S=\int\d
t\left\{\frac{1}{2}\,(\dot\phi)^2(\id)+\frac{1}{2}[\phi\,(O\,\phi)](\id)
-\sum_{j=3}^n\frac{g_j}{j!}\,(\phi^j)(\id)\right\},
\end{equation}
where $g_j$'s are constants and $O$ is a linear operator from the
group algebra to the group algebra. In a more explicit form,
\begin{align}\label{f.37}
S=\int\d t\Bigg\{\frac{1}{2}&\int\d U_1\,\d
U_2\;\left[\dot\phi(U_1)\,\dot\phi(U_2)+\int\d
U\;\phi(U_1)\,O(U_2,U)\,\phi(U)\right]\,\delta(U_1\,U_2)\cr
-&\sum_{j=3}^n\frac{g_j}{j!}\int\left[\prod_{l=1}^j\d
U_l\;\phi(U_l)\right]\,\delta(U_1\cdots U_j)\Bigg\}.
\end{align}

This action would have a symmetry under
\begin{equation}\label{f.38}
\phi(U)\to\det(U_\lambda)\,\phi(U),
\end{equation}
where $\lambda$ is a representation of the group, provided
\begin{equation}\label{f.39}
O(U_2,U)=O(U)\,\delta(U_2\,U^{-1}).
\end{equation}
From now on, it is assumed that this is the case. So
\begin{align}\label{f.40}
S=\int\d t\Bigg\{\frac{1}{2}&\int\d U_1\,\d
U_2\;\left[\dot\phi(U_1)\,\dot\phi(U_2)+
\phi(U_1)\,O(U_2)\,\phi(U_2)\right]\,\delta(U_1\,U_2)\cr
-&\sum_{j=3}^n\frac{g_j}{j!}\int\left[\prod_{l=1}^j\d
U_l\;\phi(U_l)\right]\,\delta(U_1\cdots U_j)\Bigg\},
\end{align}
A simple choice for $O$ is
\begin{equation}\label{f.41}
O(U)=c\,\chi_\lambda(U+U^{-1}-2\,\id)-m^2,
\end{equation}
where $\lambda$ is a representation of the group, and $c$ and $m$
are constants. An argument for the plausibility of this choice is
the following. Consider a Lie group and a group element near its
identity, so that
\begin{align}\label{f.42}
U_\lambda=&\exp(\tilde k^a\,T_{a\,\lambda}),\cr
\approx&\id_\lambda+\tilde k^a\,T_{a\,\lambda}+\frac{1}{2}\,
(\tilde k^a\,T_{a\,\lambda})^2,
\end{align}
where $T_a$'s are the generators of the group. One has
\begin{equation}\label{f.43}
O(U)\approx c\,\chi_\lambda(T_a\,T_b)\,\tilde k^a\,\tilde k^b-m^2,
\end{equation}
which is a constant plus a bilinear form in $\tilde{\mathbf{k}}$,
just as was expected for an ordinary scalar field. In fact, if one
introduces a small constant $\ell$ so that $\tilde k$ is
proportional to $\ell$, and $c$ is proportional to $\ell^{-2}$,
then in the limit $\ell\to 0$ the expression (\ref{f.43}) is
exactly equal to a constant plus a bilinear form.

An action of the form (\ref{f.40}) with the choice (\ref{f.41}),
has also a symmetry under
\begin{equation}\label{f.44}
\phi(U)\to\phi(V\,U\,V^{-1}),
\end{equation}
where $V$ is an arbitrary member of the group.

One can write the action (\ref{f.40}) in terms of the Fourier
transform of the field in time:
\begin{equation}\label{f.45}
\phi(t,U)=:\int\frac{\d\omega}{2\,\pi}\;\exp(-\ir\,\omega\,t)
\,\check\phi(\omega,U),
\end{equation}
to arrive at
\begin{align}\label{f.46}
S=&\frac{1}{2}\int\frac{\d\omega_1\,\d
U_1}{2\,\pi}\,\frac{\d\omega_2\,\d U_2}{2\,\pi}
\;\left[-\omega_1\,\omega_2\,\check\phi(U_1)\,\check\phi(U_2)+
\check\phi(U_1)\,O(U_2)\,\check\phi(U_2)\right]\cr
&\times[2\,\pi\,\delta(\omega_1+\omega_2)\,\delta(U_1\,U_2)]\cr
-&\sum_{j=3}^n\frac{g_j}{j!}\int\left[\prod_{l=1}^j
\frac{\d\omega_l\,\d U_l}{2\,\pi}\;\check\phi(U_l)\right]
\,[2\,\pi\,\delta(\omega_1+\cdots+\omega_j)\,\delta(U_1\cdots
U_j)].
\end{align}
The first two terms represent a free action, with the propagator
\begin{equation}\label{f.47}
\check\Delta(\omega,U):=\frac{\ir\,\hbar}{\omega^2+O(U)},
\end{equation}
while the third term contains interactions. Any Feynman graph
would consist of propagators, and $j$-line vertices to which one
assigns
\begin{equation}\label{f.48}
V_j:=\frac{g_j}{\ir\,\hbar\,j!}\sum_{\Pi}\{2\,\pi\,
\delta(\omega_1+\cdots+\omega_j)\,\delta[U_{\Pi(1)}\cdots
U_{\Pi(j)}]\},
\end{equation}
where the summation runs over all $j$-permutations. Also, for any
internal line there is an integration over $U$ and $\omega$, with
the measure $\d\omega\,\d U/(2\,\pi)$. As the group is assumed to
be compact, the integration over the group is integration over a
compact volume. Hence there would be no ultraviolet divergences.

One can compare this model to a field theory on a group manifold.
In the latter model, the integration in (\ref{f.37}) or
(\ref{f.40}) would be on the position not on the momenta, and the
operator $O$ would be the differentiation with respect to the
coordinates. In a model on a group manifold, the position
coordinates are still commuting but the momenta are not. Here the
situation is reversed, and it is not only a matter of convenience.
The operator $O$ determines which model is being investigated: it
is algebraic in terms of the momenta and the differentiation in terms
of the position. For models on group manifolds with compact
groups, there would be no infrared divergences while here there is
no ultraviolet divergence. The fact that for noncommutative
geometry based on Lie groups, the momenta are still commuting, is
the reason that here the momentum picture has been preferred to
the position picture.

One can also write the action (\ref{f.40}) in terms of the matrix
elements of the field defined like (\ref{f.22}). One arrives at
\begin{equation}\label{f.49}
S=\int\d t\sum_\lambda\,\frac{\dim_\lambda}{\vo}
\,\mathrm{tr}\left[\frac{1}{2}\left(\dot\phi_\lambda^2
+\phi_\lambda\,\tilde\phi_\lambda\right)
-\sum_{j=3}^n\frac{g_j}{j!}\,\phi_\lambda^j,\right],
\end{equation}
where $\phi_\lambda$ is defined like (\ref{f.22}), the summation
goes over irreducible representations of the group, and one has
\begin{equation}\label{f.50}
\tilde\phi_{\lambda\,a}{}^b:=\frac{\vo}{\dim_\lambda}\,
\sum_{\sigma\,\rho}\frac{\dim_\sigma}{\vo}\,\frac{\dim_\rho}{\vo}\,
C_{\lambda,\,\sigma\,\rho\,a}{}^{b\,c\,d}{}_{e\,f}
\,O_{\sigma\,c}{}^e\,\phi_{\rho\,e}{}^f,
\end{equation}
where $C$ is the kernel appearing in the decomposition of the
product of the two representations $\sigma$ and $\rho$:
\begin{equation}\label{f.51}
U_\sigma{}^c{}_e\,U_\rho{}^d{}_f=\sum_\lambda
C_{\lambda,\,\sigma\,\rho\,a}{}^{b\,c\,d}{}_{e\,f}
\,U_\lambda{}^a{}_b.
\end{equation}
Perhaps the form (\ref{f.49}) shows more clearly the role of all
representations of the group in the model, compared to models
based on a single representation.

\section{An example: the group SU(2)}
For the group SU(2), one has
\begin{equation}\label{f.52}
f^a{}_{b\,c}=\epsilon^a{}_{b\,c}.
\end{equation}
A group element $U$ can be characterized by the coordinates
$(k^1,k^2,k^3)$ such that
\begin{equation}\label{f.53}
U=\exp(\ell\,k^a\,T_a),
\end{equation}
where $\ell$ is a constant. The invariant measure is
\begin{equation}\label{f.54}
\d U=\frac{\sin^2(\ell\,k/2)}{(\ell\, k/2)^2}\,\frac{\d^3
k}{(2\,\pi)^3},
\end{equation}
where
\begin{equation}\label{f.55}
k:=\left(\delta_{a\,b}\,k^a\,k^b\right)^{1/2}.
\end{equation}
The reason for this particular choice of normalization is that for
small values of $k$, (\ref{f.54}) reduces to the integration
measure corresponding to the ordinary space. The integration
region for the coordinates is
\begin{equation}\label{f.56}
k\leq\frac{2\,\pi}{\ell}.
\end{equation}
Of course this does not mean that we are dealing with functions on a
three-dimensional ball of radius $(2\,\pi/\ell)$. The functions
are defined on a three-sphere, $S^3$. The situation is very much like the
case of functions defined on a circle. One can say that the
argument of such a function is between $0$ and $(2\,\pi)$, while
it is understood that the values of the function for $0$ and
$(2\,\pi)$ are the same.

In the small-$k$ limit, one also has
\begin{equation}\label{f.57}
\delta(U_1\,\cdots\,U_l)\approx (2\,\pi)^3\, \delta
(\mathbf{k}_1+\cdots+\mathbf{k}_l),
\end{equation}
which ensures an approximate momentum conservation. The exact
conservation law, however, is that at each vertex the product of
incoming group elements should be unity. For the case of a
three-leg vertex, one can write this condition as
\begin{equation}\label{f.58}
\exp(\ell\,k_1^a\,T_a)\,\exp(\ell\,k_2^a\,T_a)\,
\exp(\ell\,k_3^a\,T_a)=1,
\end{equation}
or a similar condition in which $\mathbf{k}_1$ is replaced by
$\mathbf{k}_2$ and vice versa. One has
\begin{equation}\label{f.59}
\exp(\ell\,k_1^a\,T_a)\,\exp(\ell\,k_2^a\,T_a)=:
\exp[\ell\,\gamma^a(\mathbf{k}_1,\mathbf{k}_2)\,T_a],
\end{equation}
where the function $\boldsymbol{\gamma}$ enjoys the properties
\begin{align}\label{f.60}
\boldsymbol{\gamma}[\mathbf{k}_1,
\boldsymbol{\gamma}(\mathbf{k}_2,\mathbf{k}_3)]=&
\boldsymbol{\gamma}[\boldsymbol{\gamma}(\mathbf{k}_1,\mathbf{k}_2),
\mathbf{k}_3],\\ \label{f.61}
\boldsymbol{\gamma}(-\mathbf{k}_1,-\mathbf{k}_2)=&
-\boldsymbol{\gamma}(\mathbf{k}_2,\mathbf{k}_1),\\ \label{f.62}
\boldsymbol{\gamma}(\mathbf{k},-\mathbf{k})=&0.
\end{align}
So that (\ref{f.58}) becomes one of the three equivalent forms
\begin{align}\label{f.63}
\mathbf{k}_3=&-\boldsymbol{\gamma}(\mathbf{k}_1,\mathbf{k}_2),\cr
\mathbf{k}_2=&-\boldsymbol{\gamma}(\mathbf{k}_3,\mathbf{k}_1),\cr
\mathbf{k}_1=&-\boldsymbol{\gamma}(\mathbf{k}_2,\mathbf{k}_3).
\end{align}
The explicit form of $\boldsymbol{\gamma}$ is obtained from
\begin{align}\label{f.64}
\cos\frac{\ell\,\gamma}{2}=&
\cos\frac{\ell\,k_1}{2}\,\cos\frac{\ell\,k_2}{2}-
\frac{\mathbf{k}_1\cdot\mathbf{k}_2}{k_1\,k_2}\,
\sin\frac{\ell\,k_1}{2}\,\sin\frac{\ell\,k_2}{2},\cr
\frac{\gamma^a}{\gamma}\,\sin\frac{\ell\,\gamma}{2}=&
\epsilon^a{}_{b\,c}\,\frac{k_1^b\,k_2^c}{k_1\,k_2}\,
\sin\frac{\ell\,k_1}{2}\,\sin\frac{\ell\,k_2}{2}\cr &+
\frac{k_1^a}{k_1}\,\sin\frac{\ell\,k_1}{2}\,\cos\frac{\ell\,k_2}{2}+
\frac{k_2^a}{k_2}\,\sin\frac{\ell\,k_2}{2}\,\cos\frac{\ell\,k_1}{2}.
\end{align}
It is easy to see that in the limit $\ell\to 0$,
$\boldsymbol{\gamma}$ tends to $\mathbf{k}_1+\mathbf{k}_2$, as
expected.

The choice (\ref{f.41}) for $O$ turns to be
\begin{equation}\label{f.65}
O=2\,c\,
\left\{\frac{\displaystyle{\sin\left[\left(s+\frac{1}{2}\right)\,\ell\,k\right]}}
{\displaystyle{\sin\frac{\ell\,k}{2}}}-(2\,s+1)\right\}-m^2,
\end{equation}
where $s$ is the spin of the representation. For small values of
$k$, this is turned to
\begin{equation}\label{f.66}
O\approx-c\,\frac{s\,(s+1)\,(2\,s+1)}{3}\,(\ell k)^2-m^2,\qquad
(\ell\,k)\ll 1.
\end{equation}
One chooses $c$ so that in the small-$k$ limit $O$ takes the
ordinary form of the propagator inverse:
\begin{equation}\label{f.67}
O\approx-k^2-m^2,\qquad (\ell\,k)\ll 1.
\end{equation}
Choosing
\begin{equation}\label{f.68}
c=\frac{3}{s\,(s+1)\,(2\,s+1)\,\ell^2},
\end{equation}
the propagator becomes
\begin{equation}\label{f.69}
\check\Delta(\omega,\tilde{\mathbf{k}})=\frac{\ir\,\hbar}
{\omega^2+\displaystyle{\frac{6}{s\,(s+1)\,(2\,s+1)\,\ell^2}}\,
\left\{\frac{\displaystyle{\sin\left[\left(s+\frac{1}{2}\right)\,
\ell\,k\right]}}
{\displaystyle{\sin\frac{\ell\,k}{2}}}-(2\,s+1)\right\}-m^2}.
\end{equation}
It is easy to see that in the limit $\ell\to 0$, the usual
commutative propagator is recovered.

Similar things holds for the group SO(3). One only has to replace
the integration region by
$$ k\leq\frac{\pi}{\ell}.\eqno{(56')}$$

\section{Concluding remarks}
A real scalar field theory was investigated constructed on a
noncommutative space, the commutation relations of which are those
of a compact Lie group. To avoid explicit calculus on such a
noncommutative space, everything was defined on the momentum
space. This space is commutative and one can attribute
well-defined (local) coordinates to it, so that ordinary
differential and integral calculus (on manifolds) can be performed
on it. As far as observables of field theories are concerned, this
momentum representation is sufficient. The Feynman rules for
perturbative field theory were obtained for the noncommutative
model, and it was seen that for small momenta these are the same
as the corresponding rules for ordinary field theories, as
expected. Another way to state this is that there is a length
parameter in the noncommutative theory so that if this length
tends to zero, one recovers the results of ordinary field
theories.

Some comments are in order. As the commutation relations for the
space coordinates are the commutation relations of the generators
of a compact lie group, say SU(2) or SO(3), the eigenvalues of the
space coordinates are discrete. Roughly speaking, such theories
resemble theories defined on lattices rather than on continua. But
generally in lattice theories the rotational symmetry is broken,
while in a noncommutative theory based on the group SO(3) this is
not the case. This similarity between the noncommutative theories
discussed here and lattice theories is directly related to the
fact that in these noncommutative theories (which are based on
compact groups) there are no ultraviolet (UV) divergences. This is
simply a result of the fact that the integration region for loop
integrations is not ${\Bbb R}^4$, but ${\Bbb R}$ times a compact
manifold. (This compact manifold is the group manifold). This
UV-finiteness of the model is reminiscent of the old expectation
that in noncommutative spaces the theory might be free from the
divergences caused by the short distance behavior of physical
quantities. In this sense noncommutative theories based on compact
groups resemble ordinary (commutative theories) with a momentum
cutoff.

It would be interesting to mention the fate of the UV/IR mixing
phenomena \cite{MRS}. As a generic property of models defined on
canonical noncommutative spaces (\ref{f.1}), certain combinations
of external momenta and noncommutativity parameter $\theta$ may
appear as a dynamical cutoff in momentum space. For example, in
two external-leg diagrams of $\phi^4$ theory, the combination
$(p\circ p)^{-1/2}$ with $p\circ p:=
(p^\mu\theta_{\mu\nu}^2p^\nu)$ acts as a cutoff, causing that the
contribution of the so-called non-planar diagram be UV-finite
\cite{MRS}. In the extreme IR limit of external momenta ($p\to
0$), this cutoff tends to infinity and the result diverges. In
such a case, in the IR limit of the theory the UV divergences of
the commutative (ordinary) theory are restored. This is the
so-called UV/IR mixing. If the noncommutative theory had been
based on a commutative theory with a momentum cutoff, there would
be no UV divergence and no UV/IR mixing.

Theories discussed here are free from UV divergences, as the
momentum space is compact. In this sense, they are based on
commutative theories with a momentum cutoff. Hence there is no UV
divergence in the original theory to be restored in some IR limit,
and there is no place for UV/IR mixing.
\\
\\
\textbf{Acknowledgement}:  This work was partially supported by
the research council of the Alzahra University.

\end{document}